\documentclass[aps,prl,twocolumn,superscriptaddress]{revtex4}
\usepackage{epsf,graphicx}
\usepackage{amssymb}
\usepackage{amsmath}
\usepackage{latexsym,bm,array,amsfonts,multirow}
\usepackage{color}
\usepackage{ulem}
\begin{document}

\title{Weak electronic correlations in the cobalt oxychalcogenide superconductor Na$_2$CoSe$_2$O}
\author{Zhenchao Wu}
\affiliation{Key Laboratory of Quantum Theory and Applications of MoE $\&$ School of Physical Science and Technology,
	Lanzhou University, Lanzhou 730000, People Republic of China}
\affiliation{Lanzhou Center for Theoretical Physics, Key Laboratory of Theoretical Physics of Gansu Province,
	Lanzhou University, Lanzhou 730000, People Republic of China}
\affiliation{Beijing National Laboratory for Condensed Matter Physics and
	Institute of Physics, Chinese Academy of Sciences, Beijing 100190, China}
\author{Yingying Cao}
\affiliation{Beijing National Laboratory for Condensed Matter Physics and
	Institute of Physics, Chinese Academy of Sciences, Beijing 100190, China}
\affiliation{University of Chinese Academy of Sciences, Beijing 100049, China}
\author{Hong-Gang Luo}
\affiliation{Key Laboratory of Quantum Theory and Applications of MoE $\&$ School of Physical Science and Technology, Lanzhou University, Lanzhou 730000, People Republic of China}
\affiliation{Lanzhou Center for Theoretical Physics, Key Laboratory of Theoretical Physics of Gansu Province, Lanzhou University, Lanzhou 730000, People Republic of China}
\author{Yi-feng Yang}
\email[]{yifeng@iphy.ac.cn}
\affiliation{Beijing National Laboratory for Condensed Matter Physics and Institute of
	Physics, Chinese Academy of Sciences, Beijing 100190, China}
\affiliation{University of Chinese Academy of Sciences, Beijing 100049, China}
\affiliation{Songshan Lake Materials Laboratory, Dongguan, Guangdong 523808, China}
\date{\today}

\begin{abstract}
Motivated by the newly discovered Co-based superconductor Na$_2$CoSe$_2$O, we performed systematic calculations of its electronic band structures using the density functional theory (DFT) plus the dynamical mean-field theory (DMFT) approaches. Our comparative studies reveal weakly correlated and itinerant nature of the Co-3$d$ electrons and show no sign of fluctuating local moments as expected in many other unconventional superconductors, although the Co $e_g$ orbitals are close to half filling. These suggest that Na$_2$CoSe$_2$O is a normal paramagnetic metal and its superconductivity might not be of strongly correlated nature, contrary to the initial speculation. We suggest future investigations of electron-phonon interactions to clarify its pairing mechanism. 
\end{abstract}
\maketitle

Exploring unconventional superconductors driven by strong electronic correlations have attracted extensive intereste during the past decades \cite{Bednorz1986, Keimer2015, Kamihara2008, Stewart2011, Steglich1979, White2015, Cao2018, Balents2020}. In the high-temperature cuprate superconductors, the electron pairing occurs primarily within the two-dimensional CuO$_2$ planes and is generally believed to arise from the superexchange interaction between Cu 3$d_{x^2-y^2}$ electrons mediated by the O $p_{x/y}$ orbitals \cite{Tacon2011}. In iron-based superconductors, the pairing is induced by spin density wave fluctuations between electron and hole pockets \cite{Mazin2008}. The recently-discovered nickel-based superconductors including the infinite-layer RNiO$_2$ \cite{Li2019Nature,Osada2020}, the bilayer La$_3$Ni$_2$O$_7$ \cite{Sun2023b,Hou2023,Zhang2023c}, and the trilayer La$_4$Ni$_3$O$_{10}$ \cite{Sakakibara2023b,Li2024a,Zhu2023,Zhang2023m,Li2024} also share some similarities and exhibit layered structures and strong electronic correlations. Their superconductivity has also been ascribed to the superexchange interaction between Ni 3$d$ electrons \cite{Wang2020, Yang2023, Qin2023, Xie2024, Chen2024, Qin2024}, possibly involving complex interplay of orbital-selective Mott, Hund, and Kondo physics \cite{Zhang2020, Cao2023}.

Co is located between Fe and Ni in the periodic table, making it a promising candidate for unconventional superconductivity. But to the best of our knowledge, the sole recognized example of Co-based unconventional superconductor is the hydrated layered cobaltate Na$_x$CoO$_2$·yH$_2$O \cite{Takada2003}. Its superconductivity emerges in close proximity to a spin-$1/2$ Mott insulating state \cite{Baskaran2003, Wilhelm2015}. Achieving high-quality samples of Na$_x$CoO$_2$·yH$_2$O remains challenging, as the superconductivity only appears in the low Na content range ($x = 0.22-0.47$) and requires water to be integrated into the crystal structure \cite{Chen2004, Ueland2004, Schaak2003}. Precise control over both Na concentration and water content is crucial for producing a single-phase Na$_x$CoO$_2$·yH$_2$O material with reliable superconducting properties. 

A recent breakthrough of Co-based unconventional superconductor is the synthesis of Na$_2$CoSe$_2$O \cite{Cheng2024}, which is the first 3$d$ transition metal oxychalcogenide superconductor with distinct structural and chemical characteristics. It also has a layered structure, with alternating layers of edge-sharing Na$_6$O octahedra and CoSe$_6$ octahedra. The Co ions form a two-dimensional triangular lattice, resembling that in Na$_x$CoO$_2$·yH$_2$O but replacing CoO$_2$ by CoSe$_2$ as the conducting layer and the insulating Na$_x$(H$_2$O)$_y$ layer by the Na$_2$O layer. Although it has a superconducting transition temperature $T_c$ only of 6.3 K, Na$_2$CoSe$_2$O exhibits a number of unusual physical properties such as quasi-two-dimensionality, low carrier concentration, multiple bands, and an extremely high upper critical field. All of these point to the possibility that the superconductivity in Na$_2$CoSe$_2$O  may be unconventional. It is therefore interesting to understand the properties of Na$_2$CoSe$_2$O from the perspective of correlated electronic structure calculations.

In this work, we employ the density functional theory (DFT) \cite{2014WIEN2k,WIEN2k} plus the dynamical mean-field theory (DMFT) \cite{Georges1996RMP, Anisimov1997JPCM, Lichtenstein1998PRB, Kotliar2006RMP, Held2008JPCM, Haule2010PRB} approaches to study the correlated band structures of Na$_2$CoSe$_2$O. The DFT band structures reveal that Na$_2$CoSe$_2$O is a negative charge transfer superconductor. However, we find no significant differences between the DFT and DFT+DMFT band structures, both showing small band renormalization and weak electronic correlations. We also observe no characteristic upper and lower Hubbard bands of Co 3$d$ orbitals. Our results raise the question concerning the true pairing mechanism in Na$_2$CoSe$_2$O and suggest future investigations of potential electron-phonon interactions in this system.

\begin{figure}[t]
	\begin{center}
		\includegraphics[width=0.48\textwidth]{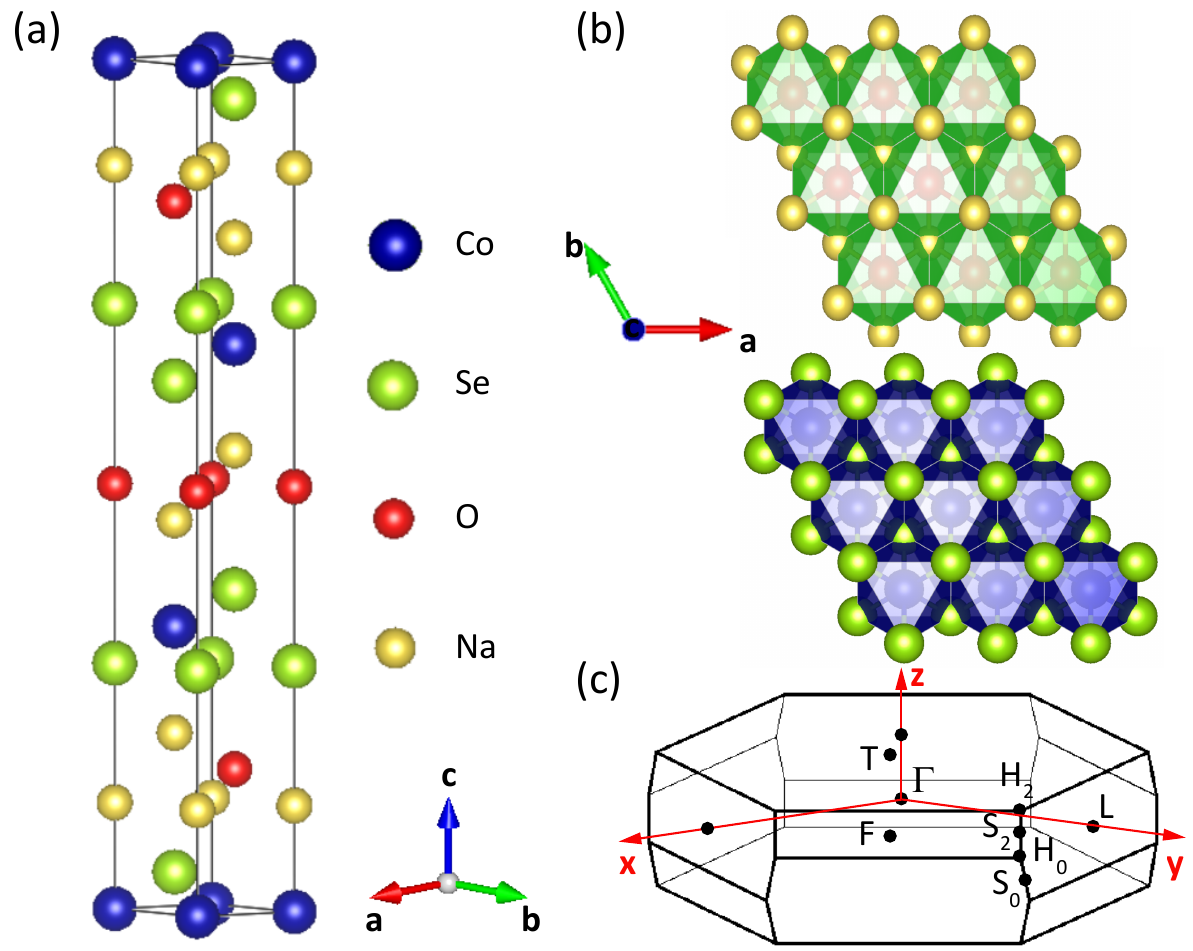}
		\caption{(a) Illustration of the crystal structure of Na$_2$CoSe$_2$O. Yellow, red, blue, and green spheres represent Na, O, Co, and Se atoms, respectively. (b) Structures of the Na$_2$O layer and the CoSe$_2$ layer. (c) The high symmetry points in the Brillouin zone used in our calculations.}
		\label{fig1}
	\end{center}
\end{figure}

Na$_2$CoSe$_2$O crystallizes in the trigonal space group R$\bar{3}$m, as shown in Fig. \ref{fig1}(a). The experimental lattice parameters are $a = 3.5161(9)$ $\text{Å}$ and $c = 28.745(11)$ $\text{Å}$. The CoSe$_2$ layers and Na$_2$O layers stack alternately along the $c$ direction. In the CoSe$_2$ layers, Co ions are arranged into a two-dimensional triangular lattice, while Se ions are located alternately above and below the plane as seen in Fig. \ref{fig1}(b), forming edge-shared CoSe$_6$ octahedra. DFT calculations were performed using the full-potential augmented plane-wave plus local orbital method and the Perdew-Burke-Ernzerhof exchange-correlation potential as implemented in the WIEN2K package \cite{2014WIEN2k, WIEN2k, Perdew1996PRL}. For the DFT+DMFT calculations, we include the local Coulomb repulsion $U = 6$ eV and the Hund's rule coupling $J = 0.8$ eV for all five Co 3$d$ orbitals and use the exact scheme for the double counting \cite{Haule2010PRB}. Other values of $U$ and $J$ have also been examined and the results remain qualitatively robust. We choose the hybridization expansion continuous-time quantum Monte Carlo (CTQMC) as the DMFT impurity solver \cite{Haule2007PRB}, and obtain the spectral function by applying the maximum entropy for the analytic continuation of the self-energies \cite{Jarrell1996PR}. Figure 1(c) shows the high symmetry points in the Brillouin zone used in the calculations.

\begin{figure}[t]
	\begin{center}
	\includegraphics[width=0.48\textwidth]{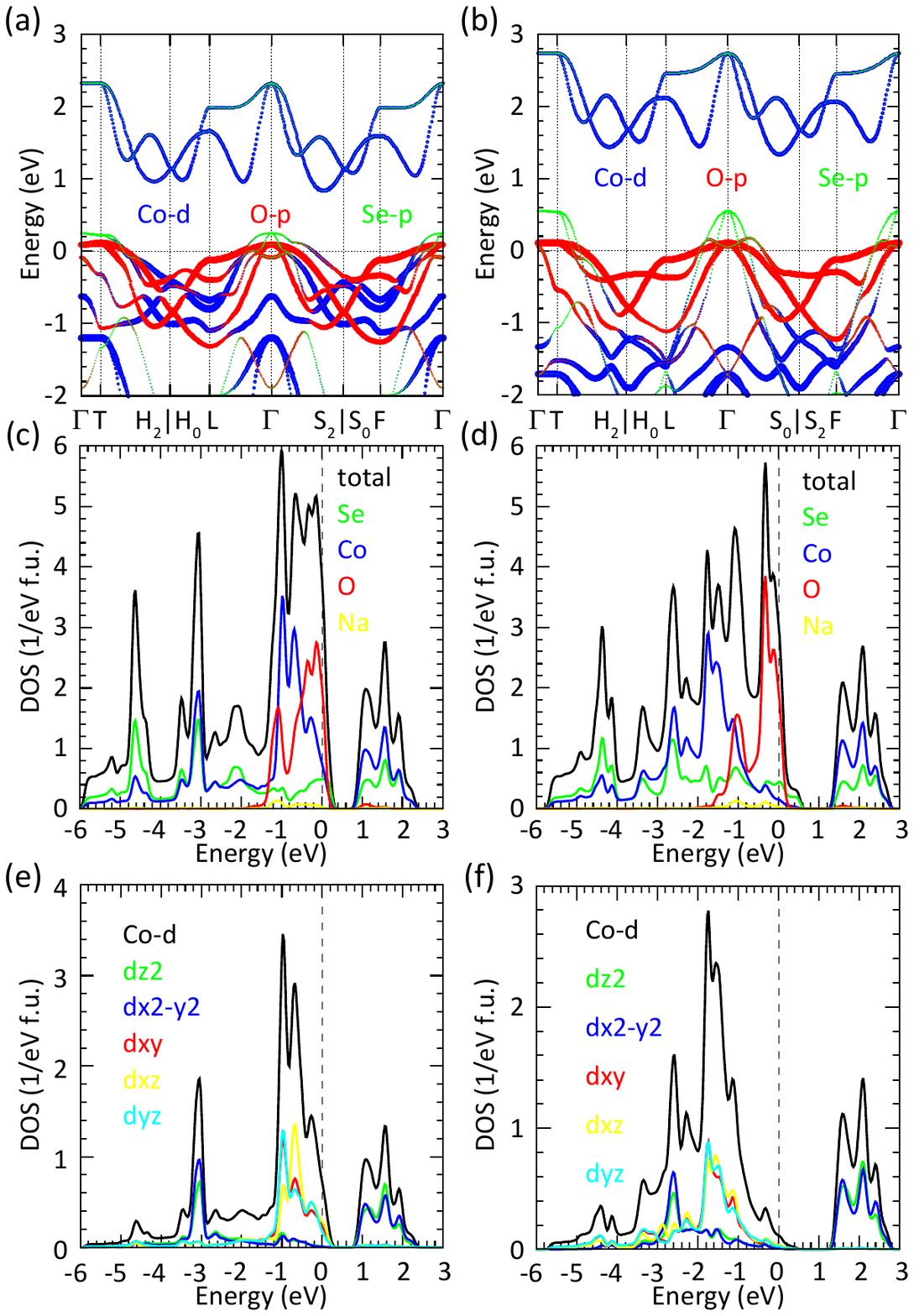}
	\caption{Atomic or orbital projected electronic band structures and densities of states (DOS) of Na$_2$CoSe$_2$O by (a)(c)(e) DFT and (b)(d)(f) DFT+$U$ with $U=4$ eV.}
	\label{fig2}	
	\end{center}
\end{figure}

The DFT band structures are plotted in Fig. \ref{fig2}. Also shown are the typical DFT+$U$ results with $U=4$ eV for non-spin-polarized calculations. The features are overall consistent, showing multiple bands crossing the Fermi energy that come mainly from the O 2$p$ and Se 4$p$ orbitals, while a small contribution from the Co 3$d$ orbitals is seen near the Fermi energy in DFT but pushed down further away by the Coulomb interaction $U$. From the band plot in Figs. \ref{fig2}(a) and \ref{fig2}(b), we expect that the Co 3$d$ orbitals are split into $t_{2g}$ and $e_g$ manifolds, with the $t_{2g}$ orbitals being almost fully occupied below the Fermi energy and the $e_{g}$ orbitals lying 1 eV well above the Fermi energy. However, the DFT occupation numbers are found to be 1.02, 1.03, 1.75, 1.75, and 1.72 for $d_{z^2}$, $d_{x^2-y^2}$, $d_{xz}$, $d_{yz}$, and $d_{xy}$, respectively, while that from DFT+$U$ are 0.91, 0.94, 1.85, 1.81, and 1.81. This indicates that while the $t_{2g}$ orbitals are indeed almost fully occupied, the $e_g$ orbitals are approximately half-occupied, a quite bizarre result compared to the band plot. To understand this, we plot the atom and orbital-resolved density of states (DOS) in Figs. \ref{fig2}(c)-(f) over a larger energy window. It turns out that the $e_g$ orbitals show spectral weights both above the Fermi energy and around -3 eV below the Fermi energy, a quite unexpected result for the paramagnetic DFT calculations. This suggests considerable band splitting of the $e_g$ orbitals, possibly attributed to the special crystal structure of edge-shared CoSe$_6$ octahedra in the CoSe$_2$ layer. As a result, the Co $d_{z^2}$ and $d_{x^2-y^2}$ orbitals between neighboring octahedra form a zigzag-shaped bond through Se, leading to a significant bonding-antibonding splitting. This is supported by the similar pronounced DOS at the peak positions of the Co $d_{z^2}$, Co $d_{x^2-y^2}$, and Se orbitals. It seems to suggest that the $e_g$ orbitals are strongly correlated and hence the superconductivity is unconventional. In both calculations with and without $U$, there are considerable amounts of holes on the O and Se orbitals, suggesting that Na$_2$CoSe$_2$O is a negative charge transfer superconductor \cite{Mizokawa1991, Bisogni2016, Rogge2018} and differs from iron-based superconductors, where the multiple bands crossing the Fermi energy primarily originate from the correlated Fe 3$d$ orbitals \cite{Lebegue2007, Paglione2010}.

\begin{figure}[t]
	\begin{center}
	\includegraphics[width=0.48\textwidth]{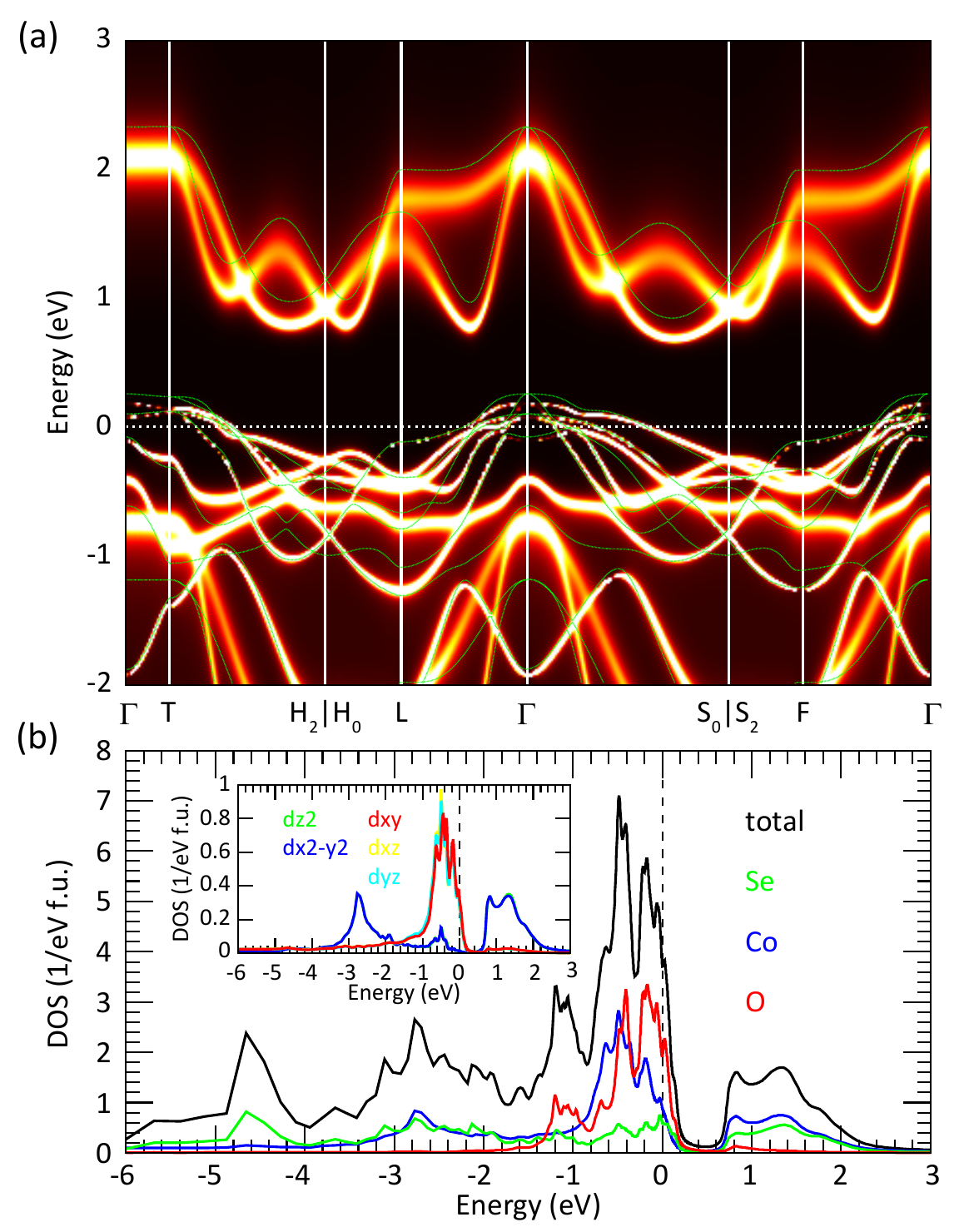}
	\caption{(a) Momentum-resolved spectral functions calculated by DFT+DMFT at $T = 100$ K for $U=6$ eV and $J=0.8$ eV. The green lines represent the bands calculated by DFT. (b) The corresponding atomic projected densities of states. The inset shows the partial densities of states of the Co-3$d$ orbitals.}
	\label{fig3}
	\end{center}
\end{figure}

To clarify whether or not the nearly half-filled $e_g$ orbitals are indeed strongly correlated, we performed DFT+DMFT calculations and plot the calculated density of states and spectral functions for $U=6$ eV and $J=0.8$ eV at $T = 100$ K in Fig. \ref{fig3}. We find no significant differences from the nonmagnetic DFT and DFT+$U$ calculations, suggesting small quasiparticle renormalization and weak electronic correlations in Na$_2$CoSe$_2$O. We also do not see any sign of lower and upper Hubbard bands in the Co 3$d$ spectral functions. The electron occupancy numbers from DMFT calculations are 0.98, 0.98, 1.78, 1.78, and 1.75 for $d_{z^2}$, $d_{x^2-y^2}$, $d_{xz}$, $d_{yz}$, and $d_{xy}$ orbitals, respectively, close to the DFT calculations. Comparing Fig. \ref{fig3} and Fig. \ref{fig2}, it is evident that DFT+DMFT yields qualitatively similar results as DFT and DFT+$U$.

\begin{figure}[t]
	\begin{center}
		\includegraphics[width=0.48\textwidth]{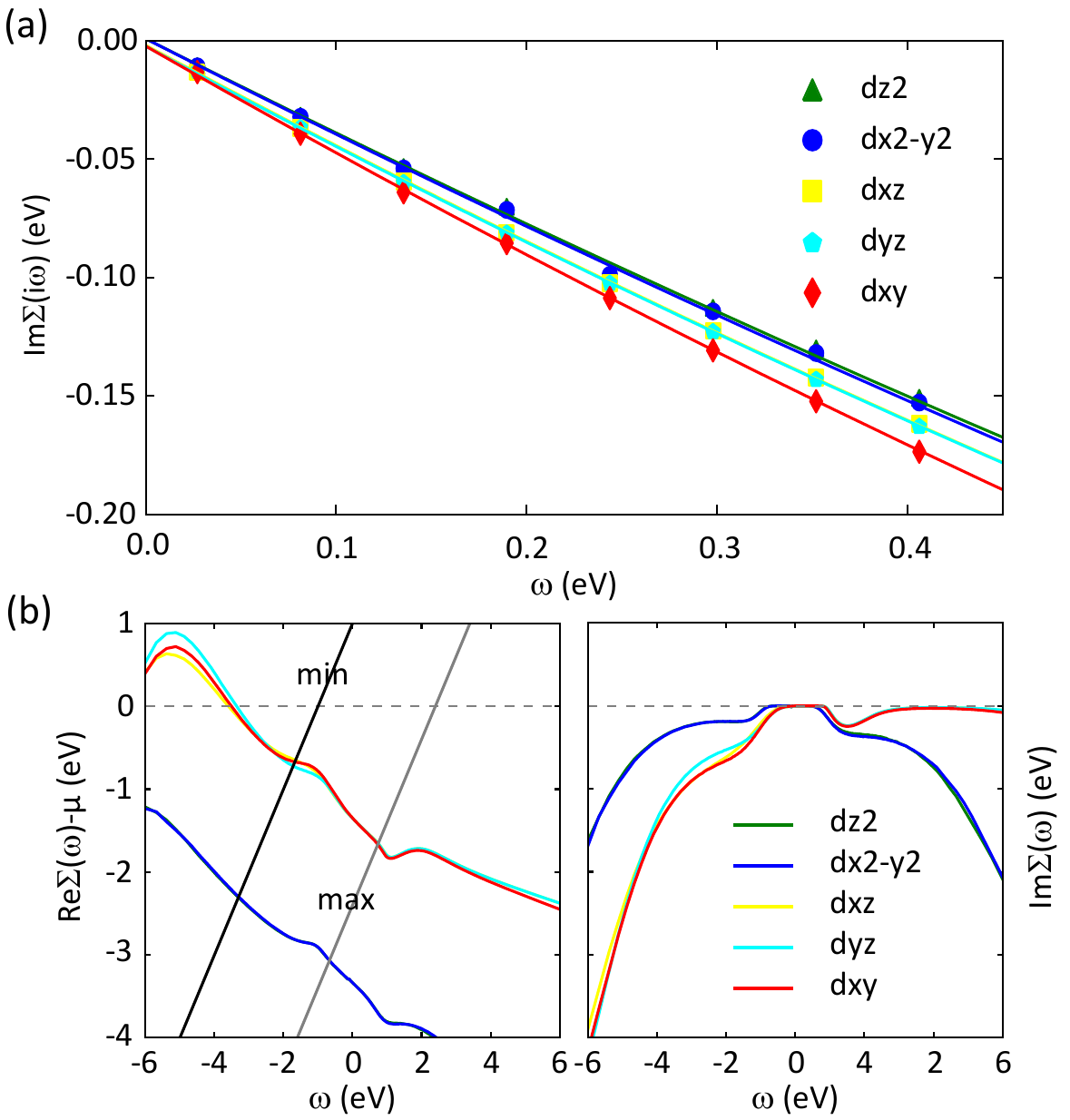}
		\caption{(a) Imaginary part of the orbital-dependent self-energies in Matsubara frequency. The solid lines are fittings to the data using the fourth-order polynomial. (b) Orbital-dependent Co 3$d$ self-energies  $\Sigma(\omega)$ in real frequency at $T = 100$ K obtained by DFT+DMFT and the maximum entropy. The notation "min"("max") stands for the minimal (maximal) energy range of the Co-3$d$ bands obtained in DFT.}
		\label{fig4}
	\end{center}
\end{figure}

We further calculated the renormalization factor or the mass enhancement defined as $m^*/m=Z^{-1}=1-\partial \mathrm{Im}\Sigma(i\omega)/\partial\omega|_{\omega\rightarrow0^+}$, where $\Sigma(i\omega)$ is the self-energy in Matsubara frequency. We obtain the exact value of the renormalization factor by fitting the lowest six points of the imaginary part of the orbital-dependent self-energy with a fourth-order polynomial \cite{Mravlje2011PRL} as shown in Fig. \ref{fig4}(a). For $U = 6$ eV and $J = 0.8$ eV, we find $m^*/m\approx$ 1.8, 1.8,1.5, 1.5, and 1.6 for the Co $d_{z^2}$, $d_{x^2-y^2}$, $d_{xz}$, $d_{yz}$, and $d_{xy}$ orbitals, respectively. These values vary weakly with $U$ and $J$. In particular, for $U = 6$ eV and $J = 0.2$ eV, we obtain $m^*/m\approx$ 1.6, 1.7, 1.6, 1.7, and 1.7, while for $U = 4$ eV and $J = 0.8$ eV, they change to 1.4, 1.4, 1.3, 1.3, and 1.3. The weak renormalization suggests that none of the Co 3$d$ orbitals in Na$_2$CoSe$_2$O are strongly correlated. As shown in Fig. \ref{fig3}(b), the DOS of all orbitals are qualitatively close to the DFT results. This supports that the splitting of the almost half-filled $e_g$ orbitals are not from the Mott physics but rather from the Co-Se-Co bond between neighboring octahedra in the CoSe$_2$ plane as discussed earlier for DFT calculations.

To further confirm that there exist no Hubbard bands for the Co 3$d$ orbitals, we performed analytic continuation for the self-energy using the maximum entropy \cite{Jarrell1996PR}. The results for the real part and imaginary part of the self-energy in real frequency are shown in Fig. \ref{fig4}(b), where the solid lines labelled as min and max stand for the energy ranges of the Co 3$d$ bands from DFT. Their intersections with the real part of the self-energies roughly give the poles of the momentum-resolved Green's function as determined by $\mathrm{det}[ \hat{H}^{\mathrm{DFT}}(\mathbf{k})+\mathrm{Re}\hat{\Sigma}(\omega)-\omega\hat{I}] =0$ \cite{Poteryaev2007}. The lower and upper Hubbard bands are manifested as nontrivial solutions of this equation. However, only a trivial intersection is seen in Fig. \ref{fig4}(a). We thus conclude that the Co 3$d$ electrons in Na$_2$CoSe$_2$O are weakly correlated and can be well described by DFT. The spectra near the Fermi energy are governed mainly by the O and Se $p$ orbitals. Thus, the Co-3$d$ orbitals in Na$_2$CoSe$_2$O may not be of strongly correlated nature.

\begin{figure}[t]
	\begin{center}
		\includegraphics[width=0.48\textwidth]{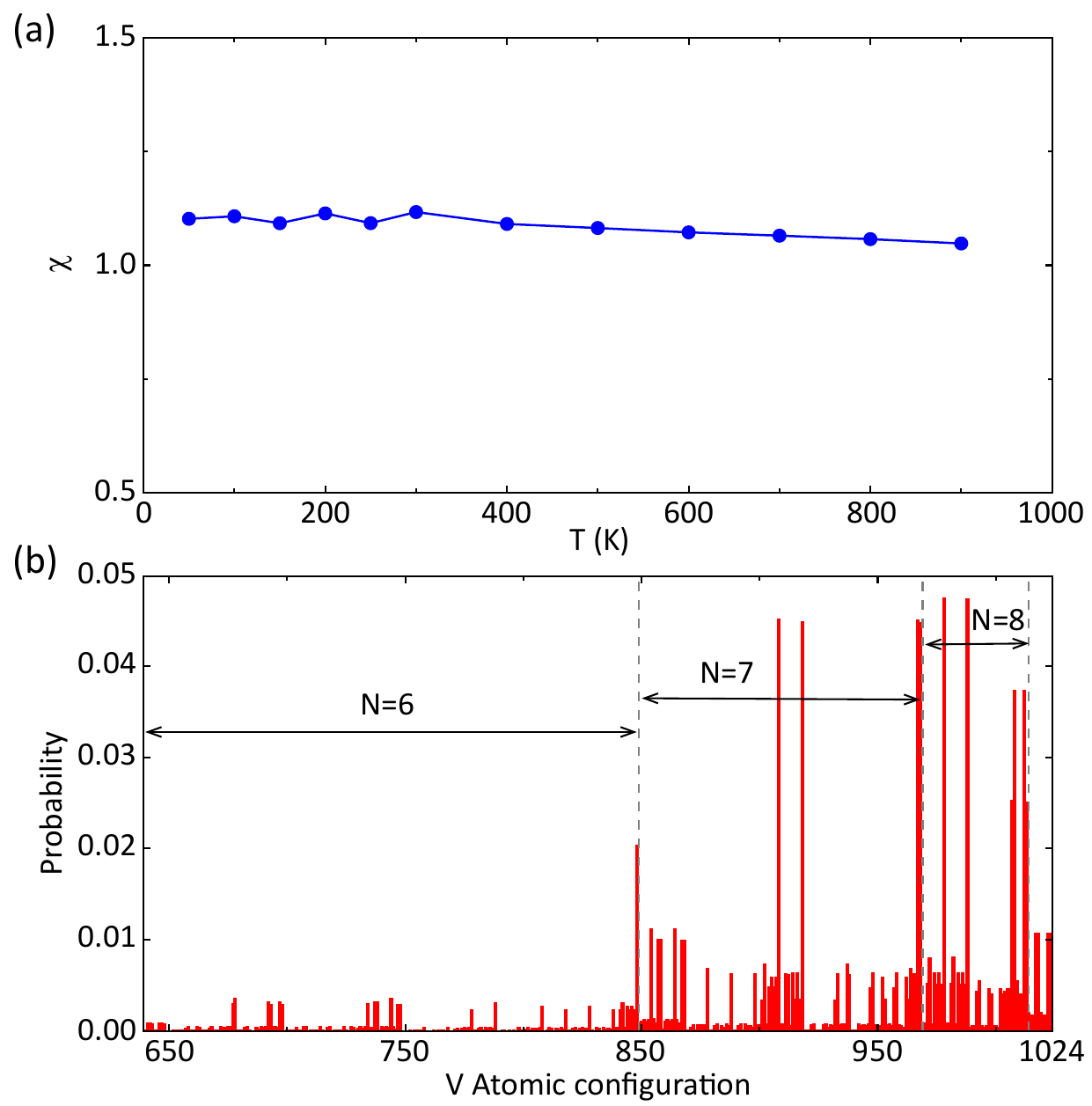}
		\caption{(a) Temperature dependence of the spin susceptibility obtained from CTQMC. (b) Probability distribution of the atomic configuration of V-3$d$ orbitals in Na$_2$CoSe$_2$O.}
		\label{fig5}	
	\end{center}
\end{figure}

Our conclusion is further supported by the calculations of the static local spin susceptibility, which is defined as $\chi=\int^{\beta}_0\left\langle S_z(\tau)S_z(0)\right\rangle d\tau$, where $\beta$ represents the inverse temperature and $\tau$ is the imaginary time. For well-defined local moments near the Mott instability, the susceptibility typically exhibits the Curie-Weiss behavior in the paramagnetic phase at high temperatures. However, as shown in Fig. \ref{fig5}(a), the susceptibility remains almost constant across the investigated temperature range from 50 K up to 900 K. This reflects characteristic Pauli paramagnetism of itinerant electrons in the absence of local moments as in good metals. To understand this, Fig. \ref{fig5}(b) shows the probability distribution of various atomic configurations in the V 3$d$ shell. Within the DFT+DMFT framework, each V ion contains a total number of 1024 configurations on 3$d$ orbitals. We find only those states with $N$ = 7, 8, 9 electrons have considerable probabilities. For each $N$, we organize the spin states in descending order of their $|S_z|$ values. The similar probabilities of many charge and spin states are consistent with weak correlations of Co-3$d$ electrons in Na$_2$CoSe$_2$O. This leads to the speculation that the superconductivity in Na$_2$CoSe$_2$O may not be of strongly correlated nature. We hence suggest future experiments or theoretical calculations of the phonons to establish this issue.

To summarize, we conduct systematic DFT+DMFT analyses of the electronic structure of the recently discovered superconductor Na$_2$CoSe$_2$O, including  the spectral function, the density of states, the renormalization factor, the susceptibility, and the atomic configurations. Our calculations show that Na$_2$CoSe$_2$O is a negative charge transfer superconductor with multiple bands crossing the Fermi level, but exhibits different multi-band characteristics compared to iron-based superconductors. Comparisons between DFT and DFT+DMFT calculations reveal small band renormalization and weak electronic correlations in Na$_2$CoSe$_2$O. Our static local spin susceptibility calculations further confirm that the Co 3$d$ electrons exhibit weakly correlated behavior with no sign of fluctuating local moments. These suggest that the superconductivity in Na$_2$CoSe$_2$O may not be of strongly correlated nature as in copper-, iron-, or nickel-based superconductors. We suggest future experimental studies or theoretical calculations of electron-phonon interactions to clarify its pairing mechanism.

This work was supported by the National Natural Science Foundation of China (Grant No. 12474136) and the Strategic Priority Research Program of the Chinese Academy of Sciences (Grant No. XDB33010100). Numerical computations were performed on Hefei advanced computing center.


\begin{thebibliography}{99}
	
	% cuprate superconductors
	\bibitem{Bednorz1986} J. G. Bednorz and K. A. Muller, Possible high	Tc superconductivity in the Ba-La-Cu-O system,	Z. Physik B - Condensed Matter \textbf{64}, 189 (1986).
	\bibitem{Keimer2015} B. Keimer, S. A. Kivelson, M. R. Norman, S. Uchida, and J. Zaanen, From quantum matter to high-temperature superconductivity in copper oxides, Nature (London) \textbf{518}, 179 (2015).
	
	% iron superconductors
	\bibitem{Kamihara2008} Y. Kamihara, T. Watanabe, M. Hirano, and H. Hosono, Iron-Based Layered Superconductor La[O$_{1-x}$F$_x$]FeAs (x = 0.05-0.12) with $T_c$ = 26 K, J. Am. Chem. Soc. \textbf{130}, 3296 (2008).
	\bibitem{Stewart2011}G. R. Stewart, Superconductivity in iron compounds, Rev. Mod.	Phys. \textbf{83}, 1589 (2011).
	
	% heavy-fermion superconductors
	\bibitem{Steglich1979} F. Steglich, J. Aarts, C. D. Bredl, W. Lieke, D. Meschede, W. Franz, and H. Schäfer, Superconductivity in the Presence of	Strong Pauli Paramagnetism: CeCu$_2$Si$_2$, Phys. Rev. Lett. \textbf{43}, 1892 (1979).
	\bibitem{White2015} B. D. White, J. D. Thompson, and M. B. Maple, Unconventional superconductivity in heavy-fermion compounds, Physica	C \textbf{514}, 246 (2015).
	
	% 2D 
	\bibitem{Cao2018} Y. Cao, V. Fatemi, S. Fang, K. Watanabe, T. Taniguchi, E. Kaxiras, and P. Jarillo-Herrero, Unconventional superconductivity in magic-angle graphene superlattices, Nature (London) \textbf{556}, 43 (2018).
	\bibitem{Balents2020} L. Balents, C. R. Dean, D. K. Efetov, and A. F. Young, Superconductivity and strong correlations in moiré flat bands, Nat. Phys. \textbf{16}, 725 (2020).
	
	\bibitem{Tacon2011} M. Le Tacon, G. Ghiringhelli, J. Chaloupka, M. M. Sala, V. Hinkov, M. W. Haverkort, M. Minola, M. Bakr, K. J. Zhou, S. Blanco-Canosa, C. Monney, Y. T. Song, G. L. Sun, C. T. Lin, G. M. De Luca, M. Salluzzo, G. Khaliullin, T. Schmitt, L. Braicovich, and B. Keimer, Intense paramagnon excitations in a large family of high-temperature superconductors, Nat. Phys. \textbf{7}, 725 (2011).
	
	\bibitem{Mazin2008} I. I. Mazin, D. J. Singh, M. D. Johannes, and M. H. Du, Unconventional superconductivity with a sign reversal in the order parameter of LaFeAsO$_{1-x}$F$_x$, Phys. Rev. Lett. \textbf{101}, 057003 (2008).
	
	%the discovery of superconductivity in infinite-layer RNiO2
	\bibitem{Li2019Nature} D. Li, K. Lee, B. Y. Wang, M. Osada, S. Crossley, H. R. Lee, Y. Cui, Y. Hikita, and H. Y. Hwang, Superconductivity in an Infinite-Layer Nickelate, Nature (London) \textbf{572}, 624 (2019).
	\bibitem{Osada2020}M. Osada, B. Y. Wang, B. H. Goodge, K. Lee, H. Yoon, K. Sakuma, D. Li, M.	Miura, L. F. Kourkoutis, and H. Y. Hwang, A Superconducting Praseodymium Nickelate with Infinite Layer Structure, Nano Lett. \textbf{20}, 5735 (2020).
	
	
	
	
	%the discovery of superconductivity in La3Ni2O7
	\bibitem{Sun2023b} H. Sun, M. Huo, X. Hu, J. Li, Z. Liu, Y. Han, L. Tang, Z. Mao, P. Yang, B. Wang, J. Cheng, D.-X. Yao, G.-M. Zhang, and M. Wang, Signatures of superconductivity near 80 K in a nickelate under high pressure, Nature (London) \textbf{621}, 493 (2023).
	\bibitem{Hou2023} J. Hou, P.-T. Yang, Z.-Y. Liu, J.-Y. Li, P.-F. Shan, L. Ma, G. Wang, N.-N. Wang, H.-Z. Guo, J.-P. Sun, Y. Uwatoko, M. Wang, G.-M. Zhang, B.-S. Wang, and J.-G. Cheng, Emergence of high-temperature superconducting phase in pressurized La$_3$Ni$_2$O$_7$ crystals, Chin. Phys. Lett. \textbf{40}, 117302 (2023).
	\bibitem{Zhang2023c} Y. Zhang, D. Su, Y. Huang, Z. Shan, H. Sun, M. Huo, K. Ye, J. Zhang, Z. Yang, Y. Xu, Y. Su, R. Li, M. Smidman, M. Wang, L. Jiao, and H. Yuan, High-temperature superconductivity with zero-resistance and strange-metal behavior in La$_{3}$Ni$_{2}$O$_{7-\delta}$, Nat. Phys. \textbf{20}, 1269 (2024).
	
	%the experimental observations of superconductivity in La4Ni3O10
	\bibitem{Sakakibara2023b} H. Sakakibara, M. Ochi, H. Nagata, Y. Ueki, H. Sakurai, R. Matsumoto, K. Terashima, K. Hirose, H. Ohta, M. Kato, Y. Takano, and K. Kuroki, Theoretical analysis on the possibility of superconductivity in the trilayer ruddlesden-popper nickelate La$_4$Ni$_3$O$_{10}$ under pressure and its experimental examination: comparison with La$_3$Ni$_2$O$_7$, Phys. Rev. B \textbf{109}, 144511 (2024).
	\bibitem{Li2024a} Q. Li, Y.-J. Zhang, Z.-N. Xiang, Y. Zhang, X. Zhu, and H.-H. Wen, Signature of superconductivity in pressurized La$_4$Ni$_3$O$_{10}$, Chin. Phys. Lett. \textbf{41}, 017401 (2024).
	\bibitem{Zhu2023}  Y. Zhu, E. Zhang, B. Pan, X. Chen, D. Peng, L. Chen, H. Ren, F. Liu, N. Li, Z. Xing, J. Han, J. Wang, D. Jia, H.	Wo, Y. Gu, Y. Gu, L. Ji, W. Wang, H. Gou, Y. Shen et al., Superconductivity in pressurized trilayer La$_4$Ni$_3$O$_{10-\delta}$ single crystals, Nature (London) \textbf{631}, 531 (2024).
	\bibitem{Zhang2023m} M. Zhang, C. Pei, X. Du, W. Hu, Y. Cao, Q. Wang, J. Wu, Y. Li, H. Liu, C. Wen, Y. Zhao, C. Li, W. Cao, S. Zhu, Q. Zhang, N. Yu, P. Cheng, L. Zhang, Z. Li, J. Zhao, Y. Chen, H. Guo, C. Wu, F. Yang, S. Yan, L. Yang, and Y. Qi, Superconductivity in trilayer nickelate La$_4$Ni$_3$O$_{10}$ under pressure, arXiv:2311.07423.
	\bibitem{Li2024} J. Li, C.-Q. Chen, C. Huang, Y. Han, M. Huo, X. Huang, P. Ma, Z. Qiu, J. Chen, X. Hu, L. Chen, T. Xie, B. Shen, H. Sun, D. Yao, and M. Wang, Structural transition, electric transport, and electronic structures in the compressed trilayer nickelate La$_4$Ni$_3$O$_{10}$, SCPMA, \textbf{67}, 117403 (2024).
	
	\bibitem{Wang2020} Z. Wang, G.-M. Zhang, Y.-F. Yang, and F.-C. Zhang, Distinct pairing symmetries of superconductivity in infinite-layer nickelates, Phys. Rev. B \textbf{102}, 220501(R) (2020).
	\bibitem{Yang2023} Y.-F. Yang, G.-M. Zhang, and F.-C. Zhang, Interlayer valence bonds and two-component theory for high-$T_c$ superconductivity	of La$_3$Ni$_2$O$_7$ under pressure, Phys. Rev. B \textbf{108}, L201108 (2023).
	\bibitem{Qin2023} Q. Qin and Y.-F. Yang, High-$T_c$ superconductivity by mobilizing local spin singlets and possible route to higher $T_c$ in pressurized La$_3$Ni$_2$O$_7$, Phys. Rev. B \textbf{108}, L140504 (2023).
	\bibitem{Xie2024} T. Xie, M. Huo, X. Ni, F. Shen, X. Huang, H. Sun, H. C. Walker, D. Adroja, D. Yu, B. Shen, L. He, K. Cao, and M. Wang, Neutron scattering studies on the high-$T_c$ superconductor La$_3$Ni$_2$O$_{7-\delta}$ at ambient pressure, Science Bulletin \textbf{69}, 3221 (2024).
	\bibitem{Chen2024} X. Chen, J. Choi, Z. Jiang, J. Mei, K. Jiang, J. Li, S. Agrestini, M. Garcia-Fernandez, H. Sun, X. Huang, D. Shen, M. Wang, J. Hu, Y. Lu, K.-J. Zhou, and D. Feng, Electronic and magnetic excitations in La$_3$Ni$_2$O$_7$. Nat Commun \textbf{15}, 9597 (2024).
	\bibitem{Qin2024} Q. Qin, J. Wang, and Y.-F. Yang, Frustrated superconductivity and intrinsic reduction of $T_c$ in trilayer nickelate, The Innovation Materials \textbf{2}, 100102 (2024). 	
	
	\bibitem{Zhang2020} G.-M. Zhang, Y.-F. Yang, and F.-C. Zhang, Self-doped Mott insulator for parent compounds of nickelate superconductors, Phys. Rev. B \textbf{101}, 020501(R) (2020). 
	\bibitem{Cao2023} Y. Cao and Y.-F. Yang, Flat bands promoted by Hund's rule coupling in the candidate double-layer high-temperature superconductor La$_3$Ni$_2$O$_7$, Phys. Rev. B \textbf{109}, L081105 (2024).
	
	%Superconductivity in NaxCoO2·yH2O
	\bibitem{Takada2003} K. Takada, H. Sakurai, E. Takayama-Muromachi,	F. Izumi, R. A. Dilanian, and T. Sasaki, Superconductivity in two-dimensional CoO$_2$ layers,
	Nature (London) \textbf{422}, 53 (2003).
	
	%spin-1/2 Mott insulating state
	\bibitem{Baskaran2003} G. Baskaran, Electronic Model for CoO$_2$ Layer Based Systems: Chiral Resonating Valence Bond Metal and Superconductivity, Phys. Rev. Lett. \textbf{91}, 097003(2003).
	\bibitem{Wilhelm2015} A. Wilhelm, F. Lechermann, H. Hafermann, M. I. Katsnelson, and A. I. Lichtenstein, From Hubbard bands to spin-polaron excitations in the doped Mott material Na$_x$CoO$_2$, Phys. Rev. B \textbf{91}, 155114 (2015).
	
	%Achieving high-quality samples of NaxCoO2·yH2O	remains challenging
	\bibitem{Chen2004} D. P. Chen, H. C. Chen, A. Maljuk, A. Kulakov, H. Zhang, P. Lemmens, and C. T. Lin, Single-crystal growth and investigation of Na$_x$CoO$_2$ and Na$_x$CoO$_2$·yH$_2$O, Phys. Rev.B \textbf{70}, 024506 (2004).
	\bibitem{Ueland2004} B. G. Ueland, P. Schiffer, R. E. Schaak, M. L. Foo, V. L. Miller, and R. J. Cava, Specific heat study of the Na$_{0.3}$CoO$_2$·1.3H$_2$O superconductor influence of the complex chemistry, Physica C \textbf{402}, 27 (2004).
	\bibitem{Schaak2003} R. E. Schaak, T. Klimczuk, M. L. Foo, and R. J. Cava, Superconductivity phase diagram of Na$_x$CoO$_2$·1.3H$_2$O, Nature (London) \textbf{424}, 527 (2003).
	
	%Superconductivity in Na$_2$CoSe$_2$O
	\bibitem{Cheng2024}J. Cheng, J. Bai, B. Ruan, P. Liu, Y. Huang, Q. Dong, Y. Huang, Y. Sun, C. Li, L. Zhang, Q. Liu, W. Zhu, Z. Ren and G. Chen, Superconductivity in a Layered Cobalt Oxychalcogenide Na$_2$CoSe$_2$O with a Triangular Lattice, J. Am. Chem. Soc \textbf{146}, 5908 (2024).
	
	%DFT
	\bibitem{2014WIEN2k} P. Blaha, K. Schwarz, G. K. H. Madsen, D. Kvasnicka and J. Luitz, WIEN2k, An Augmented Plane Wave + Local Orbitals Program for Calculating Crystal Properties (Karlheinz Schwarz, Techn. Universit{\"a}t Wien, Austria), (2001). ISBN 3-9501031-1-2.
	
	\bibitem{WIEN2k} P. Blaha, K. Schwarz, F. Tran, R. Laskowski, G. K. H. Madsen, and L. D. Marks, WIEN2k: An APW+lo Program for Calculating the Properties of Solids, J. Chem. Phys. \textbf{152}, 074101 (2020).
	
	%DMFT
	\bibitem{Georges1996RMP} A. Georges, G. Kotliar, W. Krauth, and M. J. Rozenberg, Dynamical Mean-Field Theory of Strongly Correlated Fermion Systems and the Limit of Infinite Dimensions, Rev. Mod. Phys. \textbf{68}, 13 (1996).
	
	\bibitem{Anisimov1997JPCM} V. I. Anisimov, A. I. Poteryaev, M. A. Korotin, A. O. Anokhin, and G. Kotliar, First-Principles Calculations of the Electronic Structure and Spectra of Strongly Correlated Systems: Dynamical Mean-Field Theory, J. Phys.: Condens. Matter \textbf{9}, 7359 (1997).
	
	\bibitem{Lichtenstein1998PRB} A. I. Lichtenstein and M. I. Katsnelson, Ab Initio Calculations of Quasiparticle Band Structure in Correlated Systems: LDA++ Approach, Phys. Rev. B \textbf{57}, 6884 (1998).
	
	\bibitem{Kotliar2006RMP} G. Kotliar, S. Y. Savrasov, K. Haule, V. S. Oudovenko, O. Parcollet, and C. A. Marianetti, Electronic Structure Calculations with Dynamical Mean-Field Theory, Rev. Mod. Phys. \textbf{78}, 865 (2006).
	
	\bibitem{Held2008JPCM} K. Held, O. K. Andersen, M. Feldbacher, A. Yamasaki, and Y. F. Yang, Bandstructure Meets Many-Body Theory: the LDA+DMFT Method, J. Phys.: Condens. Matter \textbf{20}, 064202 (2008).
	
	\bibitem{Haule2010PRB} K. Haule, C.-H. Yee, and K. Kim, Dynamical Mean-Field Theory within the Full-Potential Methods: Electronic Structure of {CeIrIn$_5$}, {CeCoIn$_5$}, and {CeRhIn$_5$}, Phys. Rev. B {\bf 81}, 195107 (2010).
	
	%GGA
	\bibitem{Perdew1996PRL} J. P. Perdew, K. Burke, and M. Ernzerhof, Generalized Gradient Approximation Made Simple, Phys. Rev. Lett. \textbf{77}, 3865 (1996).
	
	%CTQMC solver
	\bibitem{Haule2007PRB} K. Haule, Quantum Monte Carlo Impurity Solver for Cluster Dynamical Mean-Field Theory and Electronic Structure Calculations with Adjustable Cluster Base, Phys. Rev. B \textbf{75}, 155113 (2007).
	
	%the maximum entropy
	\bibitem{Jarrell1996PR} M. Jarrell and J. E. Gubernatis, Bayesian Inference and the Analytic Continuation of Imaginary-Time Quantum Monte Carlo Data, Phys. Rep. \textbf{269}, 133-195 (1996).
	
	%negative charge transfer 
	\bibitem{Mizokawa1991} T. Mizokawa, H. Namatame, A. Fujimori, K. Akeyama, H.	Kondoh, H. Kuroda, and N. Kosugi, Origin of the band gap in the	negative charge-transfer-energy compound NaCuO$_2$, Phys. Rev. Lett. \textbf{67}, 1638	(1991).
	
	\bibitem{Bisogni2016} V. Bisogni, S. Catalano, R. J. Green, M. Gibert, R. Scherwitzl, Y. Huang, V. N. Strocov, P. Zubko, S. Balandeh, J.-M. Triscone, G. Sawatzky, and T. Schmitt, Ground-state oxygen holes and the metalinsulator transition in the negative charge-transfer rare-earth nickelates, Nat. Commun. \textbf{7}, 13017 (2016).
	
	\bibitem{Rogge2018} P. C. Rogge, R. U. Chandrasena, A. Cammarata, R. J. Green, P. Shafer, B. M. Lefler, A. Huon, A. Arab, E. Arenholz, H. N. Lee,	T.-L. Lee, S. Nemšák, J. M. Rondinelli, A. X. Gray, and S. J.	May, Electronic structure of negative charge transfer CaFeO$_3$ across	the metal-insulator transition, Phys. Rev. Mater. \textbf{2}, 015002 (2018).
	
	%iron-based superconductors, where the multiple bands crossing the Fermi energy primarily originate from the correlated Fe 3$d$ orbitals
	\bibitem{Lebegue2007} S. Lebegue, Electronic Structure and Properties of the Fermi Surface of the Superconductor LaOFeP, Phys. Rev. B \textbf{75}, 035110 (2007).
	\bibitem{Paglione2010} J. Paglione and R. L. Greene, High-Temperature Superconductivity in Iron-Based Materials, Nat. Phys. \textbf{6}, 645 (2010).
	
	%fourth-order polynomial to get the effective mass
	\bibitem{Mravlje2011PRL} J. Mravlje, M. Aichhorn, T. Miyake, K. Haule, G. Kotliar, and A. Georges, Coherence-Incoherence Crossover and the Mass-Renormalization Puzzles in Sr$_2$RuO$_4$, Phys. Rev. Lett. \textbf{106}, 096401 (2011).
	
	%Their intersections with the real part of the self-energies roughly give the poles of the momentum-resolved Green's function
	\bibitem{Poteryaev2007} A. I. Poteryaev, J. M. Tomczak, S. Biermann, A. Georges, A. I. Lichtenstein, A. N. Rubtsov, T. Saha-Dasgupta, and O. K. Andersen, Enhanced crystal-field splitting and orbital-selective coherence induced by strong correlations in V$_2$O$_3$, Phys. Rev. B \textbf{76}, 085127 (2007).
\end{thebibliography}
\end{document}